\documentclass[
    ,final            % use final for the camera ready runs
    , sort&compress
  ]
  {aipproc}

\usepackage{amsmath}
\layoutstyle{8x11double}

\input{babarsym.tex}
\def\csbar  {\ensuremath{c\overline s}\xspace}
\def\Dsj    {\ensuremath{D_{sJ}}\xspace}
\def\Dszero {\ensuremath{D_{sJ}^*(2317)^+}\xspace}
\def\Dsone  {\ensuremath{D_{sJ}(2460)^+}\xspace}
\def\Dspi   {\ensuremath{D_s^+\pi^0}\xspace}
\def\Dsspi  {\ensuremath{D_s^{*+}\pi^0}\xspace}
\def\Dssg   {\ensuremath{D_s^{*+}\gamma}\xspace}
\def\Dsg    {\ensuremath{D_s^+\gamma}\xspace}
\def\Dsgpi  {\ensuremath{D_s^+\gamma\pi^0}\xspace}
\def\Dspipi {\ensuremath{D_s^+\pi^+\pi^-}\xspace}
\def\etal   {{\it et al}\xspace}

\begin{document}

\title{Charmed-strange Mesons Experimental Results}

\author{Jianchun Wang
}{
address={Syracuse University, Department of Physics, 
         Syracuse, NY 13244, U.S.A.}
}

%%%%%%%%%%%%%%%%%%%%%%%%%%%%%%%%%%%%%%%%%%%%
%%  abstract
%%%%%%%%%%%%%%%%%%%%%%%%%%%%%%%%%%%%%%%%%%%%

\begin{abstract}
Two new states in the charm strange sector, \Dszero and \Dsone,
have recently been discovered at \epem collider experiments. 
The new states are first observed in the dominant \Dspi and \Dsspi modes 
respectively and are very narrow. 
They are consistent with $0^+$ and $1^+$ P-wave \csbar mesons. 
The \Dsone meson is also observed in \Dsg and \Dspipi modes.
A review of the discoveries and possible explanations is given.

\end{abstract}

\maketitle

%%%%%%%%%%%%%%%%%%%%%%%%%%%%%%%%%%%%%%%%%%%%
%%  introduction
%%%%%%%%%%%%%%%%%%%%%%%%%%%%%%%%%%%%%%%%%%%%

\section{Introduction}

In a simplified picture, the charmed-strange meson \csbar (generically denoted as
\Dsj in this paper) is an atom of a massive charm quark and a light anti-strange
quark. 
The mass splitting of different states is the result of interaction of the spin angular 
momenta of the two quarks, $\vec{s_c}$ and $\vec{s_s}$, and the orbital angular momentum
$\vec{L}$ between them.
According to HQET~\cite{DeRujula:1976kk, Isgur:1991wq}, 
in the limit that the charm quark is infinitively heavy,
its spin is totally decoupled from the light degree of freedom. 
Then the spin of charm quark $\vec{s_c}$ and 
%the total angular momentum of light degrees of freedom 
$\vec{j}=\vec{L}+\vec{s_s}$ are conserved separately by strong interactions. 
This is the so-called heavy quark symmetry (HQS).

The charm quark, however, is not infinitively heavy, 
but it is heavier than the QCD scale \lqcd.
Thus taking $\vec{J}=\vec{L}+\vec{s_s}+\vec{s_c}$ as a good quantum number,
the two ground states ($L=0,\ J^P = 0^-, 1^-$) can be considered as $j=1/2$ doublets
and the four first orbital excited states ($L=1$) can be treated as $j=1/2$ doublets
($J^P=0^+, 1^+$) and $j=3/2$ doublets ($J^P=1^+, 2^+$)~\cite{Isgur:1991wq,Godfrey:1991wj}.

Before this year only four of these six states had been observed. 
All the observed ones are narrow.
The $0^-$ state, \Ds, is the lightest \Dsj meson and thus can decay 
only weakly~\cite{Chen:1983kr}.
The $1^-$ state, \Dss, was discovered in the electromagnetic radiative
mode $D_s^{*+}\to D_s^+\gamma$~\cite{Albrecht:1984ty}.
The kinematically allowed strong transition $D_s^{*+}\to D_s^+\pi^0$ is isospin suppressed, 
and has branching fraction of only $\sim$ 6\%~\cite{Gronberg:1995qp}.
The two observed $L=1$ states are $D_{s1}(2536)^+\to D^*K$,
and $D_{sJ}(2573)^+\to DK$~\cite{Albrecht:1989yi,Kubota:1994gn}.
Being members of $j=3/2$ doublets, they decay in D-wave not S-wave,
explaining their relatively narrow widths.

The two missing $L=1$ states ($0^+$ and $1^+$) were predicted by most potential 
models~\cite{Godfrey:1985xj,Zeng:1995vj,Gupta:1995mw,Ebert:1998en,DiPierro:2001uu}
to be massive enough that they would decay to $DK$ and $D^*K$, respectively, in a S-wave. 
The widths were thus expected to be very broad, $\sim$200-300 MeV.
There were, however, a few predictions that these states would have masses 
below $D^{(*)}K$ threshold that evidently were not paid much
attention~\cite{Bardeen:1994ae,Fayyazuddin:1993cc,Deandrea:2000yc}.
Effectively ``everyone'' thought that $D^{(*)}K$ were the modes to look for these 
two states and they were difficult to find due to the large width.
The recent discoveries reveal a different picture~\cite{Aubert:2003fg,
Stone:2003cu,Besson:2003cp,Abe:2003jk,Krokovny:2003zq,Aubert:2003pe}.

%%%%%%%%%%%%%%%%%%%%%%%%%%%%%%%%%%%%%%%%%%%%
%%  D_sJ(2317)
%%%%%%%%%%%%%%%%%%%%%%%%%%%%%%%%%%%%%%%%%%%%
\section{Discovery of $D_{sJ}^*(2317)^+$}

\begin{figure}
  \includegraphics[height=.20\textheight]{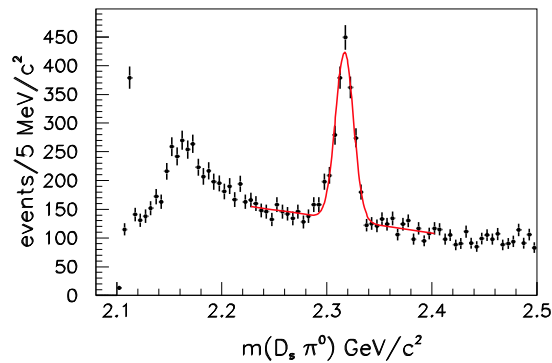}
  \caption{The \Dspi invariant mass distribution from BaBar.}
  \label{fig:babar_0p}
\end{figure}

The BaBar collaboration observed a \Dspi structure in their \epem continuum 
event sample~\cite{Aubert:2003fg}.
The center of peak is $2317.3\pm 0.4 \pm 0.8$ MeV as shown in Figure~\ref{fig:babar_0p}.
The width of the peak is $8.6 \pm 0.4$ MeV, consistent with their detector resolution.
The structure is observed in different \Ds decay modes.
It does not appear in their generic Monte Carlo simulated sample,
and thus it is not a reflection of a previously known decay.

Since the decay products of this new state must contain a charm and an 
anti-strange quark, it is natural to think that this is one of the $L=1$ 
\Dsj mesons that are still missing.
Thus it is named as \Dszero.
Furthermore, the $1^+$ meson is forbidden to decay into $0^-0^-$, 
whereas the $0^+$ meson is allowed in S-wave.
The decay angular distribution is flat after reconstruction efficiency correction,
which means either \Dszero is generated unpolarized or it is a spin-0 state.
So this new state is probably the $0^+$ \Dsj meson, though higher spin is not ruled out.

The mass of \Dszero, however, is much lighter than the $0^+$  \Dsj meson predicted
by most potential models.
For example, the model in reference~\cite{DiPierro:2001uu} worked quite well with known $D$ and \Dsj
mesons at the time it was created, and successfully predicted the mass of $0^+$ and $1^+\ D$ 
mesons that were later discovered.
It predicted the mass of $0^+$ \Dsj meson to be 2487 MeV.
The newly observed \Dszero is 170 MeV lower than the expectation,
it is even $\sim 40$ MeV below the $DK$ threshold.
And the width is much narrower ($<10$ MeV) than the prediction of $\sim$ 200-300 MeV.

%%%%%%%%%%%%%%%%%%%%%%%%%%%%%%%%%%%%%%%%%%%%
%%  D_sJ(2460)
%%%%%%%%%%%%%%%%%%%%%%%%%%%%%%%%%%%%%%%%%%%%
\section{Discovery of \Dsone}

\begin{figure}
  \includegraphics[height=.32\textheight]{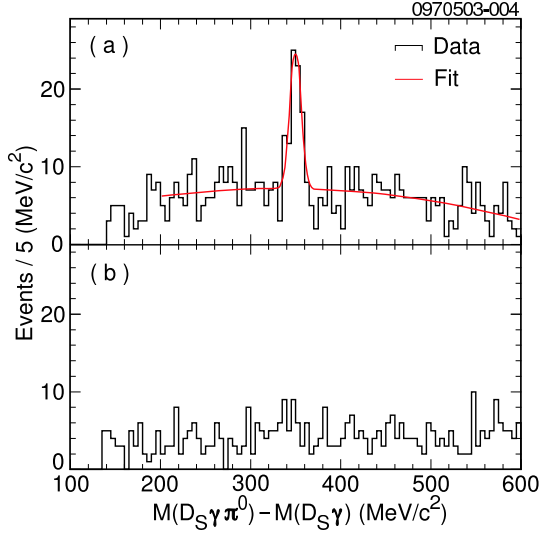}
  \caption{The \Dsgpi mass distribution from CLEO for \Dsg candidates from 
           a) \Dss signal, b) \Dss sidebands.}
  \label{fig:cleo_1p}
\end{figure}

The CLEO collaboration confirms the \Dspi resonance observed by
BaBar~\cite{Stone:2003cu,Besson:2003cp}.
They find that the measured width of the peak is $8.0_{-1.2}^{+1.3}$ MeV, 
somewhat broader than their detector resolution of $6.0\pm 0.3$ MeV.
More interestingly they also observe another state, \Dsone, at 2463 MeV that 
decays into \Dsspi (Figure~\ref{fig:cleo_1p}).

Figure~\ref{fig:cleo_1p}.a shows the invariant mass difference, 
$\Delta M = M(D_s^+\gamma\pi^0)-M(D_s^+\gamma)$.
Requiring \Dsg consistent with \Dss, they find $55\pm 10$ events in the peak. 
The center of peak is measured to be $349.8\pm 1.3$ MeV, similar to that of 
\Dszero that CLEO finds at $349.4\pm 1.0$ MeV in the 
$\Delta M = M(D_s^+\pi^0)-M(D_s^+)$ spectrum.
The width of peak is $6.1\pm 1.0$ MeV, close to the detector resolution of $6.6\pm 0.5$ MeV.
The BaBar data also shows excess in  \Dsgpi invariant mass spectrum~\cite{Aubert:2003fg},
although the conclusion reached in the publication was that further study is needed due 
to the complexity of the reflection from the \Dszero.

The $\Delta M$ values are very close for \Dszero and \Dsone.
When the \Ds from a \Dszero decay picks up a random photon,
the invariant mass of the two can fall in the selection window of \Dss.
Because of the equality of the mass difference,
when the $\pi^0$ of the same \Dszero decay is added,
the total invariant mass is consistent with \Dsone.
Thus \Dszero could reflect into \Dsone peak,
but simulation shows that this peak has width of $\sim 15$ MeV, 
much broader than the real \Dsone signal peak.
Checking the event sample from \Dss sidebands (Figure~\ref{fig:cleo_1p}.b), CLEO find that
the reflection of \Dszero could only account for 1/5 to 1/4 of events in \Dsone peak.

The reflection also exists in the opposite direction, when the single photon from \Dsone
decay is ``ignored'' and a fake \Dszero peak is created.
With the MC simulation event sample CLEO estimates the cross reflection efficiencies, 
and then extract the true number of reconstructed \Dszero and \Dsone signals.
In both peaks, about 20\% events are due to reflection.
The number of \Dsone signal is $41 \pm 12$, consistent with estimation using \Dss sidebands.

The Belle collaboration confirms both \Dszero and \Dsone states 
in continuum event sample as well as in $B$ decays that will be 
discussed in next section~\cite{Abe:2003jk,Krokovny:2003zq}.
They also observed \Dsone in \Dsg and \Dspipi modes.
After careful study of cross reflection the BaBar collaboration also confirms the 
\Dsone meson~\cite{Aubert:2003pe}.
So there is no doubt about the existence of the \Dsone state.
As \Dsone decays to $1^- 0^-$, it is most probably the missing $J^P = 1^+$ 
state decays in a S-wave.
It can not be a $0^+$ state, though other possibilities are not ruled out.
Further investigation is needed.

%%%%%%%%%%%%%%%%%%%%%%%%%%%%%%%%%%%%%%%%%%%%
%%  B decay by Belle
%%%%%%%%%%%%%%%%%%%%%%%%%%%%%%%%%%%%%%%%%%%%
\section{Observation of \Dszero and \Dsone in $B$ decays}

\begin{figure}
  \includegraphics[height=.32\textheight]{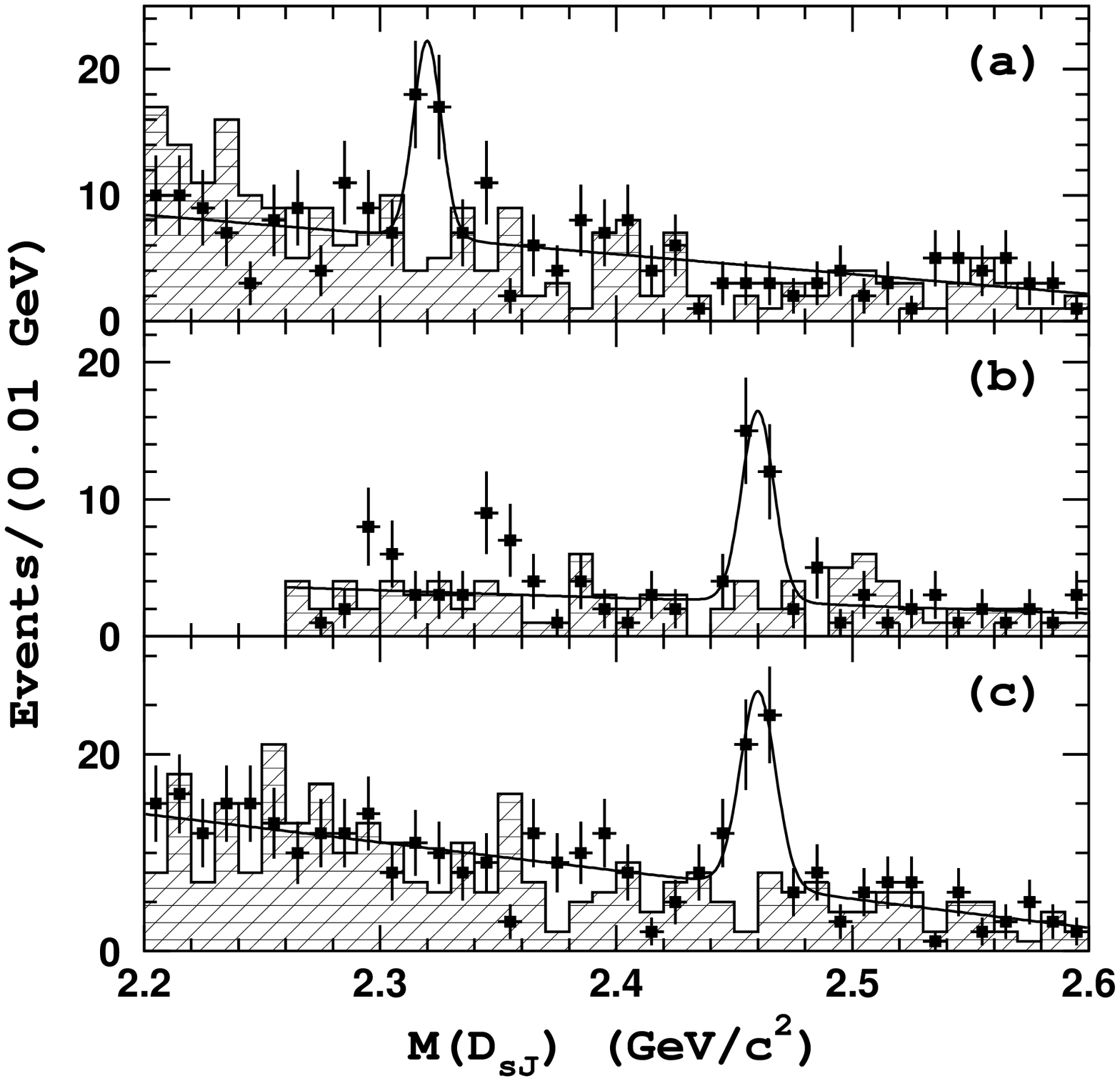}
  \caption{Invariant mass for \Dsj candidates produced in $B\to \overline{D} D_{sJ}^+$ decays for
        $D_{sJ}^+\to$ a) \Dspi, b) \Dsspi and c) \Dsg.  The hatched regions are for $\Delta E$ sidebands. }
  \label{fig:belle_m}
\end{figure}

Cross reflection of the two new \Dsj states in continuum data complicates 
the investigation. 
The cross reflection, however, is eliminated in $B$ decays as extra constraints are applied.
Belle searches for $B \to \overline{D} D_{sJ}^+$ decays of both charged
and neutral $B$~\cite{Krokovny:2003zq}.
For events whose mass and beam energy constraints are consistent with the
$B\to \overline{D} D_{sJ}^+$ decay, the invariant mass spectrum of \Dspi, 
\Dsspi and \Dsg are shown in Figure~\ref{fig:belle_m}.
Belle observes both the \Dszero and \Dsone in $B$ decays.
The peak in Figure~\ref{fig:belle_m}.c is the first observation of \Dsone $\to$ \Dsg mode.
The ratio of partial width of this mode to that of $\Dsone\to\Dsspi$ is measured
to be $0.38\pm 0.11\pm 0.04$,
consistent with $0.55 \pm 0.13 \pm 0.8$ measured in continuum data by Belle.
The branching fractions are measured to be:
\begin{align}
  \begin{split}
   {\cal B} (B \to \overline{D} \Dszero) &\times {\cal B}(\Dszero \to \Dspi) \\
                                         &= (8.5^{+2.6}_{-1.9} \pm 2.6) \times 10^{-4},\\
   {\cal B} (B \to \overline{D} \Dsone)  &\times {\cal B}(\Dsone \to \Dsspi) \\
                                         &= (17.8^{+4.5}_{-3.9} \pm 5.3) \times 10^{-4},\\
   {\cal B} (B \to \overline{D} \Dsone)  &\times {\cal B}(\Dsone \to \Dsg) \\
                                         &= (6.7^{+1.3}_{-1.2} \pm 2.0) \times 10^{-4}.
  \end{split}
  \nonumber
\end{align}

\begin{figure}
  \includegraphics[height=.28\textheight]{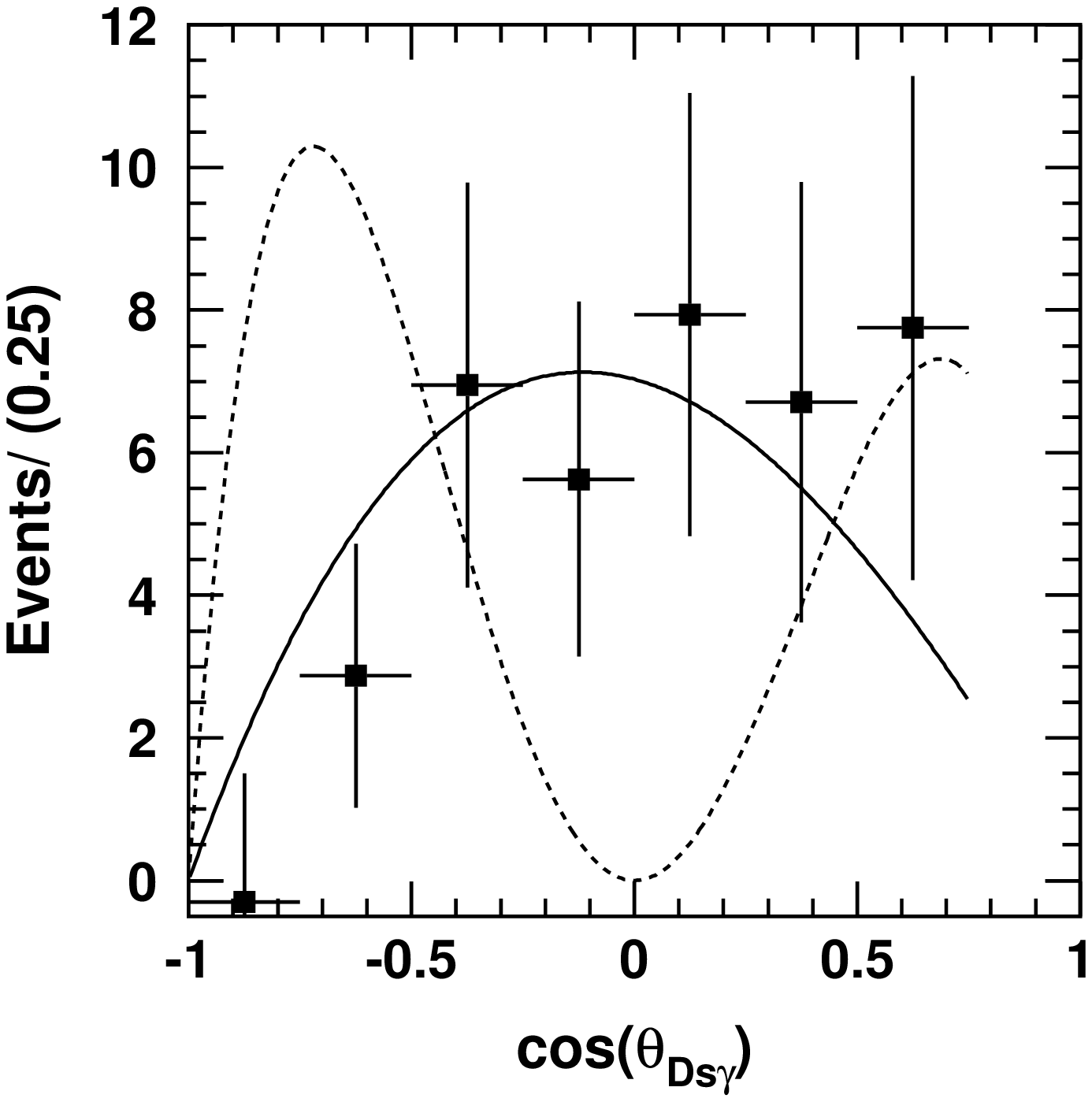}
  \caption{Helicity angular distribution of \Dsone in $B \to \overline{D} \Dsone$, \Dsone $\to$ \Dsg. The solid
           line is expectation of $1^+$ state and dotted line for $2^+$. }
  \label{fig:belle_h}
\end{figure}

The $B$ decay provides a much better laboratory to study the spin parity of the new \Dsj states.
In $B \to \overline{D} D_{sJ}^+$ decay, the \Dsj is totally longitudinally polarized as both 
$B$ and $D$ are
spin-0 particles.
Belle measures the helicity angular distribution of \Dsone in \Dsg mode shown in Figure~\ref{fig:belle_h}.
The measurement strongly supports the $1^+$ assignment.

%%%%%%%%%%%%%%%%%%%%%%%%%%%%%%%%%%%%%%%%%%%%
%%  more modes
%%%%%%%%%%%%%%%%%%%%%%%%%%%%%%%%%%%%%%%%%%%%
\section{Possible Explanation and Search of Other Decay Modes}

The world averaged mass difference are $349.1\pm 0.6$ MeV and $346.7\pm 0.8$ MeV
for \Dszero and \Dsone respectively. 
Adding the PDG value of M(\Ds) = $1968.5\pm 0.6$ MeV and M(\Dss) = $2112.4\pm 0.7$ MeV, 
the masses are  $2317.6\pm 0.8$ MeV and $2459.1\pm 1.0$ MeV.
The upper limits of width at 90\% CL are 4.6 and 5.5 MeV respectively set by 
Belle~\cite{Krokovny:2003zq}.

Since the discovery of \Dszero state several possible explanations appeared.
Cahn and Jackson use non-relativistic vector and scalar exchange forces and
recalculate within potential model to explain the mass~\cite{Cahn:2003cw}.
Van Beveran and Rupp use a unitarized meson model to explain the low mass as a 
threshold effect~\cite{vanBeveren:2003kd}.
Bardeen \etal explains that it is a normal \csbar state~\cite{Bardeen:1994ae,Bardeen:2003kt}.
Barnes \etal suggest that it is a $DK$ molecule~\cite{Barnes:2003dj}.
Several others propose different multi-quark 
models~\cite{Szczepaniak:2003vy,Cheng:2003kg,Terasaki:2003qa,Nussinov:2003uj,Datta:2003iy}.

Due to the low mass and narrow width, \Dszero has difficulty
fit in the potential models, nor does \Dsone.
They could be $DK$ and $D^*K$ molecules as they are about just 
40 MeV below the thresholds.
The mass difference between $D$ and $D^*$ is $\sim$140 MeV,
explaining the mass difference between \Dszero and \Dsone of $\sim$142 MeV.
Inside the molecule, $D^{(*)}$ and $K$ are pre-formed.
As the direct decay mode $D^{(*)}K$ is closed, quark antiquark pairs of the two have 
to be broken to form a \Ds and a $\pi^0$, thus the decay is weak.

\begin{figure}
  \includegraphics[height=.21\textheight]{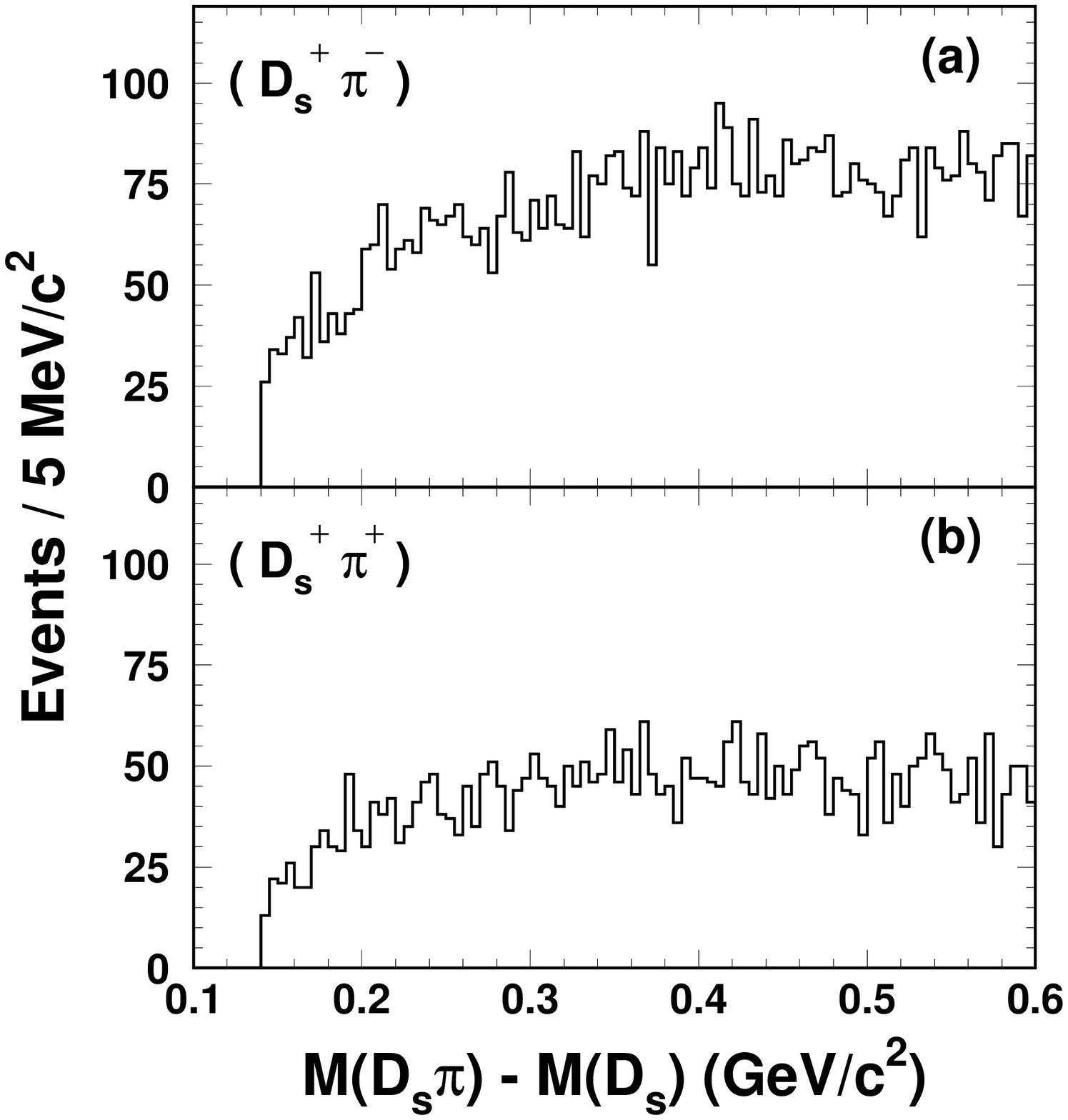}
  \hspace{-1.1cm}
  \includegraphics[height=.21\textheight]{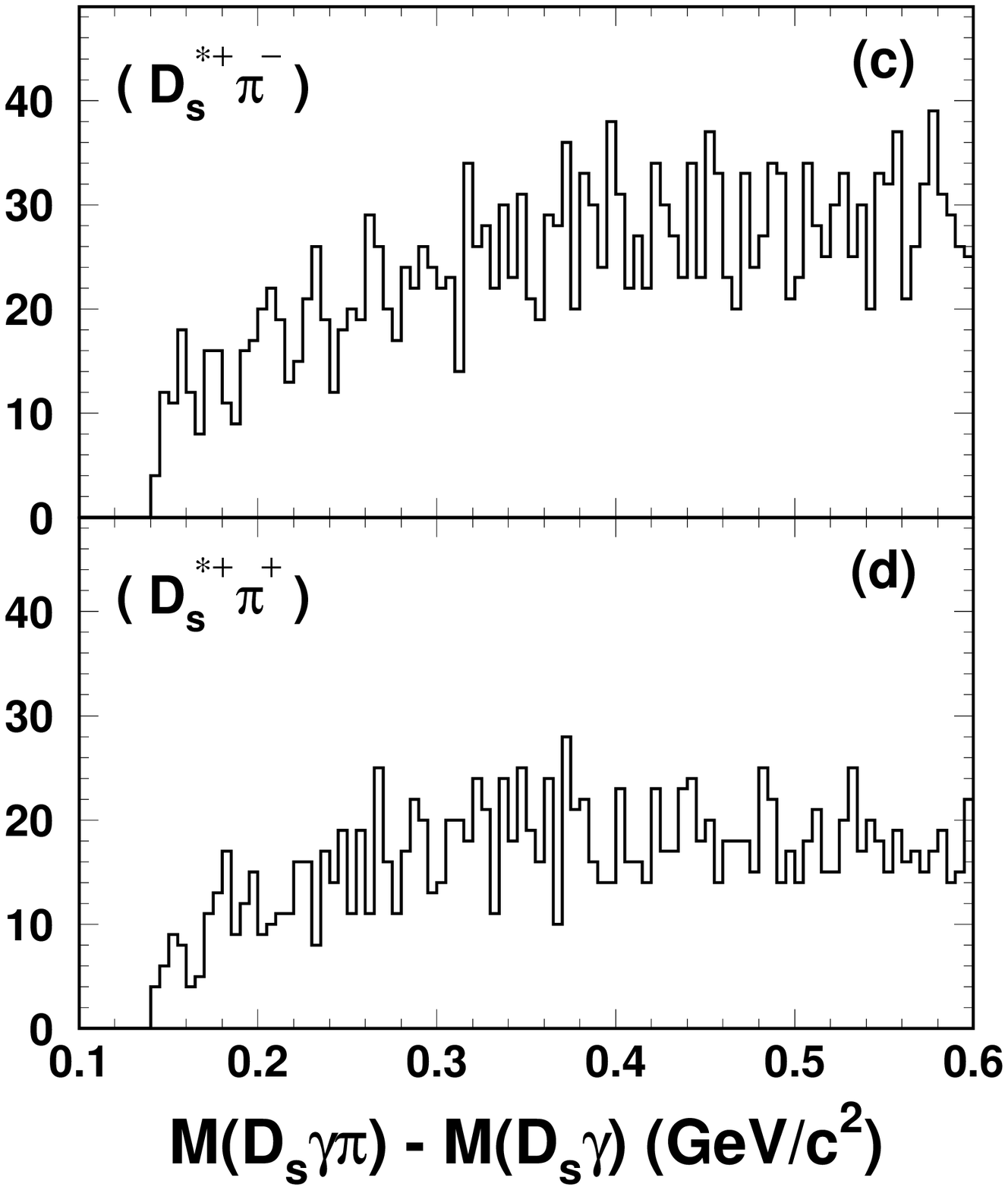}
  \caption{$D_s^{(*)+}\pi^\pm$ mass distribution from CLEO for a) opposite-signed $D_s\pi$, 
        b) same-signed $D_s\pi$, c) opposite-signed $D_s^*\pi$ and d) same-signed $D_s^*\pi$. } 
  \label{fig:cleo_dpic}
\end{figure}

The molecule picture suggests the existence of $D_s^{(*)+}\pi^\pm$ resonances. 
Observation of these resonances would strongly support molecule hypothesis as they are not conventional
\qqbar meson due to their quark content.
The CDF collaboration studies $D_s^+\pi^\pm$ modes and find no narrow structure.
%~\cite{cdf}. 
The CLEO collaboration has searched for $D_s^{(*)+}\pi^\pm$ structures as shown in Figure~\ref{fig:cleo_dpic}.
No narrow structure is found.
The productions of narrow $D_s^{(*)+}\pi^\pm$ states are at least a factor of ten lower than 
the $D_s^{(*)+}\pi^0$ modes.
This proves that \Dszero and \Dsone are iso-scalers.
It, however, does not totally rule out the molecule scenario as an iso-vector 
molecule is expected to be broad, although there is no indication of the existence
of such structure in $B$ decay sample.

The new \Dsj states fit in well the quark model as normal $0^+$ and $1^+$ \csbar mesons 
except for maybe the low masses.
Bardeen \etal couple chiral perturbation theory with a quark model representing HQET,
and in fact predicted the masses of the $0^+$ and $1^+$ \csbar mesons below $D^{(*)}K$ thresholds.
The narrow widths are due to isospin violation in the decays.
They infer that \Dszero is indeed the $0^+$ \csbar meson. 
It has an $1^+$ partner with mass splitting identical to that between $0^-$ and $1^-$ \csbar mesons,
which is backed up by the measurements.
They also calculate partial width of other decay modes as shown in Table.~\ref{tab:ratio}.
The measured ratios and limits (at 90\% C.L.) from CLEO and Belle are also listed.
The predictions are consistent with the measurements, and thus this explanation is favored.

Factorization implies that the branching fractions of $B\to\overline{D}D_{sJ}^+$ for the new \Dsj 
states be similar to that of \Ds and \Dss, which are $\sim$ 1\%.
The measurements are about a factor of ten lower.
This casts a shadow on the favored conventional \csbar explanation. 
Four-quark or molecule states, however, would have branching fraction 
consistent with the measurements~\cite{Chen:2003jp, Cheng:2003kg, Datta:2003iy}.
Browder \etal propose that these states are mixtures of \csbar and
four-quark states~\cite{Browder:2003fk}.
More experimental measurements and theoretical ideas are needed to reveal 
the true identity of these two new states.

\begin{table}
\begin{tabular}{lccc}
 \hline
   \Dszero decay & BEH & Belle & CLEO \\ \hline
   \Dspi   & $\equiv 1$ & $\equiv 1$ & $\equiv 1$ \\
   \Dspipi &    0 & $<4 \times 10^{-3}$ & $< 0.019$ \\
   \Dsg    &    0 & $< 0.05 $           & $< 0.052$ \\
   \Dsspi  &    0 &                     & $< 0.01$ \\
   \Dssg   & 0.08 & $<0.18$             & $<0.059$ \\ \hline\hline
   \Dsone decay & BEH & Belle & CLEO \\ \hline
   \Dsspi  & $\equiv 1$ & $\equiv 1$ & $\equiv 1$ \\
   \Dssg   & 0.22 & $<0.31$             & $<0.16$ \\
   \Dspi   &    0 & $<0.21$             & \\
   \Dspipi & 0.20 & $0.14\pm 0.04$ & $ < 0.08$ \\
   \Dsg    & 0.24 & $0.44\pm 0.09$ & $ < 0.49$ \\
   $D_{sJ}^*(2317)^+\gamma$
           & 0.13 &               & $<0.58$ \\ \hline
\end{tabular}
\caption{Ratio of branching fractions of different \Dszero and \Dsone modes. Limits are with 90\% CL.}
\label{tab:ratio}
\end{table}

%%%%%%%%%%%%%%%%%%%%%%%%%%%%%%%%%%%%%%%%%%%%%%%%
%% BACKMATTER
%%%%%%%%%%%%%%%%%%%%%%%%%%%%%%%%%%%%%%%%%%%%%%%%
\begin{theacknowledgments}
 The author would like to thank all those, especially Prof Sheldon Stone for
helping prepare this note.
\end{theacknowledgments}

%% If the bibliography is produced without BibTeX comment out the

\bibliographystyle{aipproc}   % if natbib is available

%\bibliography{wang}

%  \typeout{******************************************}
%  \typeout{** Please run "bibtex \jobname" to optain}
%  \typeout{** the bibliography and then re-run LaTeX}
%  \typeout{** twice to fix the references!}
%  \typeout{******************************************}

\end{document}